\magnification = \magstep1
\baselineskip = 24 true pt
\hsize = 16 truecm
\vsize = 22 truecm
\centerline {\bf {DUALITY IN QUANTUM LIOUVILLE THEORY}}
\bigskip
\centerline {L. O'Raifeartaigh, J. M. Pawlowski, and V. V. Sreedhar}
\centerline {School of Theoretical Physics}
\centerline {Dublin Institute for Advanced Studies}
\centerline {10 Burlington Road, Dublin 4}
\centerline {Ireland}
\vskip 1.5 truecm
\centerline  {\bf Abstract}

The quantisation of the two-dimensional Liouville field theory is investigated 
using the path integral, on the sphere, in the large radius limit. The general 
form of the $N$-point functions of vertex operators is found and the 
three-point function is derived explicitly. In previous work it was inferred 
that the three-point function should possess a two-dimensional lattice of poles 
in the parameter space (as opposed to a one-dimensional lattice one would 
expect from the standard Liouville potential). Here we argue that the 
two-dimensionality of the lattice has its origin in the duality of the quantum
mechanical Liouville states and we incorporate this duality into the path 
integral by using a two-exponential potential. Contrary to what one might 
expect, this does not violate conformal invariance; and has the great advantage
of producing the two-dimensional lattice in a natural way.  
\bigskip 
\bigskip
\noindent PACS: 11.10Kk, 11.10Lm, 11.25Hf, 11.25Pm, 04.60Kz 

\noindent {\it Keywords}: Liouville Theory, Path Integral, Duality  
\bigskip
\bigskip
\vfil\eject
\centerline {\bf {I. INTRODUCTION}}
\bigskip
The classical Liouville theory is defined by a real scalar field with an 
exponential potential [1]. In two dimensions it is both conformally 
invariant and integrable, and is of considerable importance in a variety of
physical problems. The quantised theory is of similar importance, particularly
in the context of string theory [2]. Accordingly, it is of interest to compute 
correlation functions in the quantised Liouville theory. As is well known, 
the scalar field itself is not a primary field with respect to the 
conformal group, and hence correlation functions are traditionally defined as 
expectation values of a set of exponential (vertex) functions which are primary
fields. However, the computation of these correlation functions turns out to be
rather tricky. In contrast to the Wess-Zumino-Witten theory (from which the 
Liouville theory can obtained by imposing a set of linear first class 
constraints on the Kac-Moody currents [3]), for example, there is no known
closed expression for the four-point function; and although the spacetime 
dependence of the two-point and three-point functions is dictated, as usual,
by conformal invariance, the computation of the their coefficients as functions
of the parameters of the theory has turned out to be quite difficult. It has 
been shown that crossing symmetry relations, for a special set of four-point
functions which are known in terms of hypergeometric functions, determine the  
coefficient of the three-point function uniquely [4]; and a form which
satisfies these requirements has been proposed by Dorn and Otto and by A. and 
Al. Zamolodchikov [5]. We shall refer to this proposal as the DOZZ proposal.
A surprising feature of the proposal is that the dependence on the 
parameters exhibits a two-dimensional lattice of poles, rather than the 
one-dimensional lattice that one would expect from a single exponential
potential.  The appearance of the two-dimensional lattice corresponds to an
unexpected duality symmetry in the quantum theory that is not present in the 
classical theory. Since this symmetry disappears in the classical limit, it 
is the opposite of the situation normally encountered with an anomaly. 
As such, the physical interpretation of the additional set 
of poles is unclear, and the residues of the proposed three-point function at
these poles can not be immediately identified within the theory.  

Motivated by the above state of affairs, we wish to consider the computation 
of the correlation functions in this paper, using the path-integral approach,
and taking a point of view that differs slightly from the conventional one. 
The difference is based on the fact that, whereas the classical Liouville 
theory admits one primary vertex field for each conformal weight, the quantum 
Liouville theory admits two such fields. In particular, it admits two distinct
exponential potentials which are conformally invariant. Accordingly, we take the
view that, for the Action in the path integral, the natural generalisation of 
the classical Liouville potential is not a single exponential potential, but a
linear combination of two independent exponential potentials whose parameters 
are arranged so as to guarantee conformal invariance.  We compute the 
three-point function in this generalised theory and as we shall see, this 
hypothesis gives a straightforward explanation of the two-dimensional lattice 
of poles and the quantum mechanical duality mentioned above. It is shown that 
our expression for the three-point function reduces to the DOZZ proposal when 
the dimensional parameters in the theory (the coefficients of the potential 
terms) obey a certain duality condition. Our approach also allows us to give a
simple expression for the four-point function. A spin-off of our approach is 
that many of our formulae are valid for any theory with two exponential 
potentials, like the Sinh-Gordon and Sine-Gordon theories; although it is only
when the parameters are related by conformal invariance, that we can carry out 
explicit computations. It will be assumed that the underlying manifold on which
the theory is defined is a compact Riemann surface with the topology of a 
two-dimensional sphere although, for computational purposes, we shall take the
infinite-volume limit.  

The paper is organised as follows. In Section II we present the arguments for
the two-exponential potential, write down the appropriate path integral, and 
specify the conditions on the parameters that make it conformally invariant.

In Section III we consider the path integral for the Liouville theory. As a 
consequence of the spherical topology of the base space, the  Liouville 
field can be separated into a constant zero mode and a fluctuating component. 
The Liouville potential has the special property that the
integration over the zero mode essentially decouples from the rest of the 
integral. By means of a Sommerfeld-Watson transform [6], the fluctuating part 
of the path integral can then be brought into a form that resembles the 
functional integral of a free scalar field theory with insertions of powers of
vertex functions -- except that the powers are not necessarily positive 
integers.  However, the result for positive integers, obtained earlier by 
Dotsenko and Fateev [7], is such that the answer for the general case can be 
obtained by an extrapolation. In Appendix B we express the Dotsenko-Fateev 
result in a form which is amenable for this extrapolation.   
The functional integration requires the regularisation
of the relevant Green's function as discussed in Appendix A, and we discuss in 
detail how this regularisation affects the Weyl invariance, and hence the 
conformal invariance of the theory. We conclude Section III by passing to the 
infinite volume limit in which we consider the translational invariance and 
scale covariance of the relevant functional integral in conformal coordinates.
We also study the dependence of the $N$-point functions on the dimensional 
parameters and show how the parameters are renormalised.  
 
In Section IV we study the covariance property of the path integral with 
respect to SL(2, C) transformations. We use this property to simplify the the 
expressions for the N-point functions. 

In Section V we consider the three-point function. It is shown that the
usual power law dependence of the three-point function can be obtained using
the SL(2, C) covariance properties of the previous section, and turns out to 
have the standard form dictated by conformal invariance. Unlike the discussions
in [5] where a new set of poles is discovered in the three point function when
it is analytically continued in the parameter space, our computations naturally
produce the full set of poles because duality is built into our construction
from the beginning.   

In Section VI we show that there is no unambiguous way to define the $N$-point  
functions for $N\leq 2$. Thus the path integral should be viewed as a sort of 
a distribution that takes meaningful values only when it is tested against at 
least three vertex functions. However, we present a proposal for the two-point 
function.   

In Section VII we show that the SL(2, C) covariance of the correlation functions
is also sufficient to fix the four-point function. Apart from the usual power
law dependence on the coordinate differences dictated by conformal invariance,
this depends on one conformally invariant cross-ratio. The function of this 
cross-ratio -- the so-called conformal block -- is a polynomial if any one of 
the vertex functions has a positive integer power.  In the generic case, where 
no power is a positive integer, one can only obtain asymptotic expansions in 
powers of the cross-ratio and its inverse. The coefficients of various powers 
in this series can be computed explicitly in terms of known three-point 
functions. 

In Section VIII, we present our conclusions.  
\bigskip 
\centerline {\bf {II. THE MODEL}}
\bigskip
The classical Euclidean Liouville Action for a real scalar field $\tilde \phi$,
on a compact manifold, which we choose to be topologically equivalent to the 
two-sphere, is given by  
$$S = \int d^2x\sqrt {g(x)}\bigl\lbrack {{1\over 4\pi}}\tilde\phi\Delta
\tilde\phi  +  {q\over 2\pi}R\tilde\phi  +  V_b (\tilde\phi)  
\bigr\rbrack\eqno(2.1)$$
where $\sqrt {g (x)}$ is the determinant of the background metric, $q$ and 
$b$ are dimensionless parameters, and $R$ is the Ricci scalar. The 
Laplace-Beltrami operator $\Delta$ and the Liouville
potential $V_b$ are given by $$\Delta = -{1\over\sqrt g}\partial_\mu\sqrt g 
g^{\mu\nu}\partial_\nu~~~{\hbox {and}}~~~ 
V_b(\tilde \phi) = \mu_b e^{2b\tilde\phi} \eqno(2.2)$$ 
respectively, $\mu_b$ being a constant parameter which has the dimensions of 
mass squared to compensate for the dimensions of $d^2x$ in the Action. 
The classical energy momentum tensor, defined as 
$T_{\mu\nu} = \delta S/\delta g^{\mu\nu}$ is
$$2\pi T_{\mu\nu} = (\partial_\mu\tilde\phi )(\partial_\nu\tilde\phi ) - 
{1\over 2}g_{\mu\nu}(\partial\tilde\phi )^2 + {1\over 2} g_{\mu\nu} V_b  
-   q (g_{\mu\nu}\Delta - \partial_\mu\partial_\nu )\tilde\phi \eqno(2.3)$$
The condition for Weyl invariance, which implies conformal invariance in the
flat space limit, is that the trace of $T_{\mu\nu}$ be proportional to $R$ on
the mass shell. This relates the parameters $q$ and $b$ by the condition
$qb = 1$. In conformal coordinates $g_{\mu\nu} (x) = e^{\sigma (x)}
\eta_{\mu\nu}$, where $\eta_{\mu\nu}$ is the flat Euclidean metric,  this 
implies that, on the mass shell, the only non-vanishing components of the 
energy-momentum tensor are $$T_{\pm} = (\partial_\pm \tilde\phi )^2 \pm q
(\partial_\pm^2\tilde\phi ) \eqno(2.4)$$   
As is well-known, the field $\tilde\phi$ is not primary with respect to the 
Virasoro algebra generated by $T_\pm$, but the fields $e^{2\alpha\tilde\phi}$ 
are primary fields of weight $(\alpha q, \alpha q)$. From this it follows that 
the only local potential that is allowed by conformal invariance {\it i.e.} has
conformal weight (1,1), is of the exponential type shown in (2.2).  

In the quantum theory, the situation is rather different because of normal
ordering and the replacement of Poisson brackets by commutators. The 
quantum analogue of the condition  
that $T_{~\mu}^\mu$ be proportional to $R$ on the mass shell, namely 
that $<T_{~\mu}^\mu >$ be proportional to $R$,  leads to a new
relation between the parameters $b$ and $q$, namely $ qb = 1 + \hbar b^2$.
The fields $e^{2\alpha\tilde\phi}$ remain primary, but have conformal
weights $(\Delta_\alpha , \Delta_\alpha )$ where
$$\Delta_\alpha  = \alpha (q - \hbar \alpha )\eqno(2.5) $$
Although the latter equation differs from its classical counterpart only by
a quantum correction, it nevertheless changes the structure of the theory
fundamentally because it means that there are now two fields for each 
conformal weight $\Delta_\alpha$, namely 
$$e^{\alpha_\pm\tilde\phi (x)} ~~~{\hbox {where}}~~~\alpha_\pm =
{q \pm \sqrt {q^2 - 4\hbar\Delta_\alpha}\over 2 \hbar}\eqno(2.6) $$
These fields are dual in the sense that 
$$\alpha_+\alpha_- = {\Delta_\alpha\over \hbar}\eqno(2.7)$$
Note that in the classical limit, $\alpha_-$ reduces to the classical
value $\alpha q$, while $\alpha_+ \rightarrow\infty$. The existence of 
dual primary 
fields, of definite conformal weight, raises the question as to whether these 
fields should be regarded as distinct or identical. From the point of view of
the Virasoro algebra, there would be no problem in identifying them because the
first class constraint
$$e^{2\alpha_+ \tilde\phi } - e^{2\alpha_- \tilde\phi } =  0\eqno(2.8)$$   
commutes with the Virasoro generators. However, this constraint does not
commute with other primary fields such as the canonical momentum
$\pi_{\tilde\phi}$. Furthermore, from the point of view of the functional
integral, it would restrict the range of the integration variable in a
non-linear way. Accordingly it would seem more reasonable not to make the
identification of $e^{\alpha_\pm\tilde\phi}$ but to treat them as distinct,
dual fields. This is the point of view we shall adopt. 

We now wish to consider the path integral formulation of the theory and 
because of the possibility of having two distinct fields of conformal
weight (1, 1) in the quantum theory, we propose to start from a path integral
with an Action which contains two exponential potentials 
$$Z = \int d\tilde\phi~~e^{-\int d^2x\sqrt {g(x)}\bigl\lbrack {1\over 4\pi}
\tilde\phi\Delta \tilde\phi  + {q\over 2\pi} R\tilde\phi +  V_b(\tilde\phi) + 
V_c(\tilde\phi) \bigr\rbrack}\eqno(2.9)$$
leaving the parameters $b~ {\hbox {and}}~ c$ to be determined by conformal
invariance. We remark in passing that an advantage of using the general
Action (2.9) is that, until we impose conformal invariance, our equations are
valid for any theory with a potential which is a sum of two different
exponentials. This includes, in particular, the Sinh-Gordon (and by 
analytic continuation, the Sine-Gordon) theory. At first sight, the proposal 
to use two potentials may seem rather radical but we shall see that it is 
perfectly compatible with conformal invariance when $b$ and $c$ are suitably 
related [8].  In fact, within the context of the path integral itself it can 
be shown (see Section III) that when renormalisation is taken into account, 
Weyl invariance, which implies conformal invariance in the flat space limit,  
requires that 
$$1 - bq + \hbar b^2 = 0~~~{\hbox{and}}~~~1 - cq + \hbar c^2 = 0 \eqno(2.10)$$ 
These equations for $b$ and $c$ are just the conditions that the potentials
$V_b$ and $V_c$ have conformal weight (1,1). If we eliminate $q$ from (2.10),
we obtain a direct relationship between $b$ and $c$
$$ (b-c)(1 - \hbar bc ) = 0 \eqno(2.11)$$ 
This equation, which has no reference to the background metric, is actually 
the necessary and sufficient condition for conformal invariance in the flat 
space limit and distinguishes the Liouville theory from other two-exponential 
theories.  

For $\hbar \neq 0$, there are obviously two solutions to equation (2.11). The 
solution $b = c$ corresponds to the case of a single potential, and can be 
recovered from the more general case $\hbar bc = 1$ by setting one of the 
dimensional parameters $\mu_b$ or $\mu_c$ equal to zero. 
We shall 
therefore consider the more general case $\hbar bc = 1$.  From now on we shall
normalise $\hbar = 1$ and thus we shall consider the path integral (2.9) with
the condition that   
$$bc = 1, ~~~q = b + c\eqno(2.12) $$
For simplicity of notation, however, we shall continue to use both $b$ and $c$
with the relationship $bc = 1$ understood. As we shall see, an important 
consequence of using the two dual potentials of conformal weight (1,1) in the 
Action is that they automatically produce the dual set of poles (in the 
parameter space) of the three point structure functions whose existence 
was inferred indirectly by other authors [5]. 

We conclude this section by defining the $N$-point function of vertex 
functions to be
$${\cal G}_N (x_I, \alpha_I)= <\prod_{I= 1}^N e^{2\alpha_I\tilde\phi(x_I)}>
=\int d\tilde\phi~ e^{-S + 2\sum_{I =1}^N\alpha_I\tilde\phi(x_I)}\eqno(2.13)$$
where $S$ is the Euclidean Action in the path integral (2.9). 
\bigskip
\centerline {\bf {III. PATH INTEGRATION, SYMMETRIES, AND RENORMALISATION}} 
\bigskip
\noindent ${\underline {The~Path~Integration}}$: It is well-known that, on a 
compact space which we choose to be topologically equivalent to the  
two-sphere, the Laplace-Beltrami operator has only one zero mode, namely the 
constant function $\phi_0$. We therefore split the field $\tilde\phi$ into its 
zero mode and its orthogonal complement $\phi$.
$$\tilde\phi (x) = \phi_0 + \phi (x),~~~  
\Delta\phi_0 = 0, \qquad \int d^2x\sqrt {g(x)}\phi(x)=0 \eqno(3.1)$$
The expression for the $N$-point function (2.13) then becomes  
$$\eqalign {{\cal G}_N (x_I, \alpha_I ) = \int d\phi_0~e^{-2\xi\phi_0}&\int 
d\phi~ e^{- U_b(\phi )e^{2b\phi_0}} e^{- U_c(\phi )e^{2c\phi_0}} \cr
&\times e^{-\int d^2x\sqrt {g(x)}\bigl\lbrack {1\over 4\pi}\phi\Delta
\phi  + {q\over 2\pi} R\phi\bigr\rbrack + 2\alpha_I\phi (x_I)}}    \eqno(3.2)$$
where
$$\xi = q - \Sigma ,~~~~\Sigma = \sum_{I = 1}^N\alpha_I~~~{\hbox {and}}~~~
U_b = \int d^2x\sqrt {g(x)} V_b(\phi )\eqno(3.3)$$  
In arriving at the above equation we have used $\int d^2x\sqrt g R = 2\pi\chi$
where $\chi$ is the Euler characteristic.  For a space which is topologically 
equivalent to the two-sphere, $\chi = 2$. Note that (2.12) implies that $b$ 
and $c$ have the same sign and because of the $\phi$-integration there is 
no loss of generality in assuming that both of them are positive. Then, for the 
zero mode integration to converge, $\xi$ must be negative. 
We will assume this to be the case. 

At first sight it would seem natural to make the Coulomb gas expansion  
$$e^{-U_b (\phi)e^{2b\phi_0}} = \sum_m {1\over m!} \Bigl\lbrack -U_b (\phi )
e^{2b\phi_0}\Bigr\rbrack^m \eqno(3.4)$$
and similarly for $b\rightarrow c$ in (3.2). The problem of computing 
correlation 
functions of arbitrary vertex operators in the Liouville theory then reduces
to the problem of computing a double infinite series of correlation functions 
of powers of integrated vertex operators in a free scalar theory. This is
reminiscent of the classical equivalence of the Liouville theory and the free
theory by a canonical transformation [9]. However, this is a little too 
restrictive because the zero mode integration then forces certain combinations
of the parameters to be positive integers. It is more convenient to  
introduce the Sommerfeld-Watson transform [6] for the exponential function 
$$ e^t =  {1\over 2\pi i}\int_{-\infty}^\infty du {(-t)^{iu}\over
\Gamma ( 1 + iu) }{\pi\over {\hbox {sinh}} \pi u}\eqno(3.5)$$
The Sommerfeld-Watson transform replaces the double series by a double
integral representation which has the virtue of manifestly displaying the
pole-like singularities (in the parameter space) of the correlation 
functions.\footnote{${}^1$}{This may be contrasted with the Coulomb gas method 
in which the exponential is expanded in powers of $U(\phi )$, in which case, 
the singularities in the individual terms are not apparent, although the series
diverges.} 
The integration in equation (3.5) is along the real axis. The above equation 
is easily verified by closing the contour of integration in the upper  
half plane or the lower half plane to enclose the poles of sinh$\pi u$
which lie along the imaginary axis at integer values and using Cauchy's 
residue theorem. For either choice of the contour, care should be taken to 
enclose the origin.  Substituting (3.5) in (3.2) and continuing the parameter
$\xi$ to $i\xi$, we get 
$$\eqalign {{\cal G}_N (x_I, \alpha_I ) &= {-{1\over 4}}\int du dv 
d\phi_0~{e^{-2i(\xi - bu - cv)\phi_0} \over {\hbox {sinh}} \pi u {\hbox {sinh}}
\pi v} {\cal R}_N (b, c; \alpha_I; iu, iv)\cr &= -{1\over 8}\int
du  dv ~{\delta (bu + cv  - \xi)\over {\hbox {sinh}} \pi u {\hbox {sinh}}\pi v}
{\cal R}_N (b, c; \alpha_I; iu, iv) }\eqno (3.6)$$  
where 
$${\cal R}_N(b, c ; \alpha_I; iu, iv) = \int d\phi{U_b^{iu}\over\Gamma (1 + iu)}
{U_c^{iv} \over \Gamma (1 + iv)}e^{-\int d^2x\sqrt {g(x)}\bigl\lbrack 
{1\over 4\pi}\phi\Delta \phi + {q\over 2\pi} R\phi \bigr\rbrack +
2\alpha_I\phi (x_I)} \eqno(3.7)$$
The function ${\cal R}_N(b, c; \alpha_I; iu, iv)$ cannot be computed directly 
for arbitrary $iu~ {\hbox {and}}~iv$. However, under certain conditions to be 
discussed later, it can be considered as an extrapolation of 
${\cal R}_N(b, c; \alpha_I; m, n)$ where $m~{\hbox {and}}~n$ are positive 
integers. For positive integers $m$ and $n$, ${\cal R}_N$ 
can be explicitly computed because it is simply a correlation function of 
integrated vertex operators in a free theory of the form   
$${\cal R}_N (b,c;\alpha_I; m, n )  = {\mu^m_b\mu^n_c\over m!n!}\int  
\prod_{i = 1}^m d^2x_i\sqrt {g (x_i)} \prod_{r = 1}^n d^2y_r\sqrt {g (y_r)}
\int d\phi e^{-\int d^2x\sqrt g  ( \phi {\Delta\over 4\pi}\phi 
+  j\phi )}    \eqno(3.8)$$ 
where 
$$j (x) = {q\over 2\pi}R (x) - \sum_{I = 1}^N {2\alpha_I\over\sqrt {g (x)}}
\delta^2 (x - z_I) - \sum_{i = 1}^m {2 b\over \sqrt {g (x)}}\delta^2 (x - x_i) 
- \sum_{r = 1}^n {2c\over \sqrt {g (x)}}\delta^2 (x - y_r)\eqno(3.9)$$ 
Of course, (3.8) mimics the Coulomb gas expansion. But the point is that 
$m$ and $n$ have to be extrapolated to the values $iu$ and $iv$ used in (3.6). 
Evaluating the Gaussian integral in (3.8) in the usual way we get 
$$\eqalign {{\cal R}_N (b, c;\alpha_I; m, n) = &{\mu_b^m\mu_c^n\over m!n!}
{1\over \sqrt {{\hbox {det}'} {\Delta\over 4\pi}}}\int \prod_{i = 1}^m d^2x_i
\sqrt {g (x_i)} \prod_{r = 1}^n d^2y_r\sqrt {g (y_r)}\cr& ~~~\times 
e^{\int d^2x\sqrt {g(x)} \int d^2y\sqrt {g(y)} j(x)G(x, y) j(y)}}\eqno(3.10)$$ 
where the prime in ${\hbox {det}}'$ means that the zero mode is omitted and 
$G(x, y)$ is the finite volume Green's function defined in Appendix A.   
As in the above equation, the subscripts of $x$, $y$, and $\alpha$ always 
run from $1$ to $m$, $1$ to $n$, and $1$ to $N$ respectively; unless otherwise 
stated. With this in mind, we shall not display the ranges of these 
subscripts from now on for the sake of notational simplicity.     

Substituting for $j$ from (3.9) we then have 
$$\eqalign{{\cal R}_N(b, c; \alpha_I; m, n) =& 
{e^{{q^2\over 4\pi^2}\int d^2x \sqrt {g(x)} d^2y \sqrt {g(y)} R(x) G (x, y)
R (y)} \over \sqrt {{\hbox {det}'} {\Delta\over 4\pi}}}
\cr&\times {\mu_b^m\mu_c^n\over m!n!} e^{\sum_{I\neq J = 1}^N 4\alpha_I\alpha_J
G(z_I, z_J)}e^{-2q\sum\alpha_Ip(z_I)} \times I_N}\eqno(3.11)$$ 
where 
$$p(x) = {1\over \pi}\int d^2y\sqrt {g(y)}~ G(x, y)R(y)\eqno(3.12)$$
$$\eqalign{I_N(b,c;\alpha_I; m, n) = \int\prod_i d^2x_i\sqrt {g (x_i)}
& e^{-2bqp(x_i)} \int \prod_r  d^2y_r\sqrt {g (y_r)}e^{-2cqp(y_r)} \cr&\times 
e^{\bigl\lbrack F_b (x_i, x) + F_c ( y_r, y) + 8 G(x_i, y_r)\bigr\rbrack }}
\eqno(3.13)$$
and 
$$ F_b(x_i, x ) =  8b\sum_I \alpha_I G(x_i , z_I) + 4b^2\sum_j G( x_i , 
x_j) \eqno(3.14)$$ 
and similarly for $F_c (y_r, y)$.  
The numerator of the first term in (3.11) will be recognised as the Polyakov
term. It implies that the centre of the Virasoro algebra is $1 + 3q^2$, where 
the one comes from the Weyl anomaly for a single real scalar field. 
This term and the denominator $\sqrt{\bigl(\hbox{det}' (\Delta/4\pi )\bigr)}$ 
of the first term in (3.11) play no further role,
so in order to simplify the calculations we drop these two terms from now on. 
We also use conformal coordinates in which $p (x)$ reduces to 
${1\over 2}{\hbox {ln}}\sqrt {g(x)}$. Since the Green's function becomes
singular when the arguments coincide, we have to renormalise it. This we do in
the standard manner discussed in Appendix A, the end result of which is that
$G(x_i, x_i)$ is taken to be ${1\over 4}{\hbox {ln}}\sqrt {g (x_i)}$. Thus
after renormalisation, integral (3.13) may be written in conformal
coordinates as 
$$I_N(b,c;\alpha_I;m,n) = \int \prod_i  d^2x_i\sqrt {g (x_i)}^{W_b} \prod_r
d^2y_r\sqrt {g (y_r)}^{W_c} e^{\bigl\lbrack F^R_b (x_i, x) + F_c^R ( y_r, y) + 
8G(x_i, y_r) \bigr\rbrack }\eqno(3.15)$$
where 
$$W_b = 1 - qb + b^2~~~{\hbox {and}}~~~W_c = 1 - qc + c^2\eqno(3.16)$$
and it is understood that the terms with coincident arguments $i = j$ and 
$r = s  $ are to be omitted in $F_R$s -- the renormalised $F$s.  
\bigskip
\noindent ${\underline {Weyl~Invariance}}$: Making a Weyl transformation 
$\sqrt g\rightarrow \lambda\sqrt g$, we see that $I_N \rightarrow 
\lambda^{mW_b + nW_c} I_N$. This factor can be absorbed in the dimensional 
parameters $\mu_b$ and $\mu_c$ in (3.11) by the scaling $\mu_b \rightarrow 
\lambda^{-W_b} \mu_b$ and $\mu_c \rightarrow \lambda^{-W_c}\mu_c$. Hence on 
retracing the steps (3.2) to (3.16) we see that a Weyl scaling of the original 
integral (3.2) has the simple effect that 
$$V_b \rightarrow \lambda^{-W_b}V_b~~~ {\hbox {and}}~~~ 
V_c \rightarrow \lambda^{-W_c}V_c \eqno(3.17)$$ 
Thus, as stated in the Introduction, Weyl invariance implies $W_b = W_c = 0$, 
which is the same as (2.10) when $\hbar $ is restored. Having extracted the 
Weyl condition, we may now consider the infinite-volume limit in which the 
Green's function takes the standard form 
$$G_0 (x, y) = -{1\over 2} {\hbox {ln}}{\mid x - y \mid\over L}\eqno(3.18)$$
$L$ being a dimensional cut-off.  
Then the equation for $I_N(b,c;\alpha_I; m,n)$ in (3.15) simplifies to 
$$\eqalign{I_N(b,c;\alpha_I;m,n) &= \int\prod_i d^2x_i\sqrt {g(x_i)}^{W_b} 
\prod_r d^2y_r \sqrt {g (y_r)}^{W_c}\cr& ~~~\times e^{\bigl\lbrack F^R_b 
(x_i, x) + F^R_c (y_r, y) - 4 {\hbox {ln}} {\mid x_i - y_r\mid\over L}\bigr
\rbrack}} \eqno(3.19)$$
where 
$$F^R_b (x_i, x) = -4b\sum_I\alpha_I {\hbox {ln}} {\mid x_i - z_I\mid\over 
L} - 2b^2 \sum_{i\neq j}{\hbox {ln}}{\mid x_i - x_j\mid\over L}\eqno(3.20) $$
This equation is actually valid for the $N$-point functions of any theory 
with two exponential potentials. What distinguishes the conformally invariant 
Liouville theory is that $W_b = W_c  = 0$ in accordance with the Weyl 
condition. In particular $bc = 1$. In that case, the integral simplifies 
further to  
$$I_N(b,c;\alpha_I;m, n)= \int dK (x_i, y_r) \prod_{i = 1}^m\prod_{r = 1}^n
\prod_{I = 1}^N \mid x_i - z_I \mid^{-4b\alpha_I} \mid y_r - z_I\mid^{-4c
\alpha_I}\eqno(3.21a)$$ 
where the measure $dK(x_i, y_r)$ is given by  
$$dK(x_i, y_r ) =  \prod_{i < j}^m \prod_{r < s}^n d^2x_id^2y_r~ \mid x_i - x_j
\mid^{-4b^2} \mid y_r - y_s \mid^{-4c^2}\mid x_i - y_r\mid^{-4}\eqno(3.21b)$$
and
$$mb + nc = \xi = q - \Sigma\eqno(3.22)$$ 
which is the zero mode constraint for integer values of $iu$ and $iv$.  
It is integral (3.21) with which we shall deal in the rest of the paper.
We shall now comment on its symmetries. 
\bigskip
\noindent ${\underline {Translational~Invariance~and~Scale~Covariance}}$:  
Integral (3.21) is obviously translationally invariant. Under a scaling, 
$z_I\rightarrow\lambda z_I$, it is easy to check, using (3.22),  that 
the measure $dK(x_i, y_r)$ transforms as 
$$dK(x_i , y_r ) \rightarrow \mid\lambda\mid^{2(mb + nc) [q - (mb + nc)]}
dK(x_i , y_r)\eqno(3.23)$$
The cross-terms between the external variables $z_I$ and the integrated 
variables $x_i$ and $y_r$ in (3.21) produce a factor $\mid\lambda\mid^{-4(mb + 
nc)\Sigma}$. Putting these two results together and using (3.22) we find that 
the integral $I$ is covariant in the sense that 
$$I_N\rightarrow I_N' = ~\mid\lambda\mid^{-2(q-\Sigma )\Sigma}I_N\eqno(3.24)$$
We should emphasise that the scaling property above takes this neat form 
only because of the constraint due to the zero mode integration. 
\bigskip
\noindent ${\underline {Renormalisation}}$: Let us consider the dependence 
of the $N$-point function on the dimensional parameters. The $\mu_b$ and 
$\mu_c$ dependence is easily read off from (3.6) and (3.8) to be of the form 
$\mu_b^m\mu_c^n$. The dependence on the cut-off $L$ can be calculated
in a straightforward manner from (3.10) by noting that each Green's function 
produces one $L$. Thus the total contribution is given by 
$e^{2(\Sigma + mb + nc)^2 {\hbox {ln}}L}$ which, upon using 
(3.22) reduces to $L^{2q^2}$. The third dimensional parameter in the 
theory is the diffeomorphic invariant short-distance cut-off $ds$ introduced 
in Appendix A. It is easy to see that since there is one subtraction to be 
made for each of the diagonal points, the total contribution is given by 
$e^{-2( mb^2 + nc^2 + \sum_I \alpha_I^2 ){\hbox {ln}}ds}$. After a 
little algebra it is easy to see, using (3.22) and the Weyl conditions 
$b^2 + 1 = qb$ and $c^2 + 1 = qc$,  that this contribution reduces to 
$(ds)^{2( m+n -q^2 + \sum_I \Delta_I )}$. Finally  let $z_0$ be a fiducial 
point from which the distances of the positions of the external variables 
$z_I$ are measured. If $L_0$ be the unit length in which these distances 
are measured, then the dependence of the $N$-point function on this unit 
length is  obtained by putting together the contributions coming 
from the $G(z_I, z_J)$ terms in (3.11) and the scaling property of the $I_N$
integral given in (3.24) and works out to be $\mid L_0\mid^{-2\sum_I\Delta_I}$.
Putting all the above factors together we find that the dimensionality of 
the $N$-point function is given by 
$$ \bigl({L\over ds}\bigr)^{2q^2} \bigl(\mu_b (ds)^2\bigr)^m\bigl(\mu_c(ds)^2
\bigr)^n \bigl({\mid L_0\mid\over ds}
\bigr)^{-2\sum_I\Delta_I}\eqno(3.25)$$ 
The first term above may be eliminated by 
absorbing it in the overall normalisation since it is independent of 
$\alpha_I$. The $ds$ in the other terms may be eliminated by renormalising 
the bare coupling constants $\mu_b$ and $\mu_c$, and $L_0$.
Thus the parametric dependence of the path integral may be written in 
dimensionless variables as  
$$\Bigl({\mu_b^R\over \mu}\Bigr)^m\Bigl({\mu_c^R\over\mu}\Bigr)^n\eqno(3.26)$$ 
where $\mu$, the renormalisation scale, has dimensions of mass squared, 
and $\mu_b^R$ and $\mu_c^R$ are the renormalised couplings.  
\bigskip
\vfil\eject
\centerline {{\bf IV. SL(2, C) COVARIANCE OF} $I_N (b,c;\alpha_I; m,n)$}  
\bigskip
Under an SL(2, C) transformation\footnote{${}^2$}{The SL(2,C) parameters 
$a,b,c,d$ should not be confused with the parameters $b$ and $c$ in the rest 
of the paper.}  
$$x_i \rightarrow  {ax_i + b\over cx_i + d},~~~~ ad-bc=1\eqno(4.1)$$ 
we have 
$$ dx_i \rightarrow {dx_i\over (cx_i + d)^2 } \quad \hbox{and} \quad 
x_i - x_j  \rightarrow  {x_i - x_j \over (cx_i + d)(cx_j + d)}\eqno(4.2)$$
It follows that the measure $dK(x_i, y_r)$ transforms as shown below:
$$dK(x_i, y_r) \rightarrow dK(x_i, y_r)\mid cx_i + d\mid^{4b (mb + nc -q)}
\mid cy_r + d\mid^{4c (mb + nc - q)}\eqno(4.3)$$
The cross-terms between the external variables and the integration variables 
produce a product of factors of the form  $\mid cx_i + d\mid^{4b\Sigma}\mid 
cy_r + d\mid^{4c\Sigma}$ for the integrated variables and $\mid cz_I + d\mid^{
4(mb + nc)\alpha_I}$ for the external variables. Putting everything together 
and using (3.22) once again, we find that all the factors corresponding to the 
integration variables cancel and $I_N$ transforms covariantly i.e.  
$$I_N (z_I)\rightarrow I_N (z_I') = I_N (z_I) \prod_{I = 1}^N \Bigl\lbrack
{1\over \mid cz_I + d\mid }\Bigr\rbrack^{-4(q-\Sigma )\alpha_I} \eqno(4.4)$$ 
It follows that the partition function is SL(2, C) invariant and the one-point 
and two-point functions are invariant with respect to two and one parameter 
non-compact sub-groups of SL(2, C) respectively. Thus, in principle, these  
functions are infinite. For the three and higher point functions, on the other 
hand, we can use the SL(2,C) covariance to simplify the integral.  To do this 
we let 
$$z_1' = {az_1 + b\over cz_1 + d} = \zeta , \qquad 
z_2' = {az_2 + b\over cz_2 + d} = 0, \qquad 
z_3' = {az_3 + b\over cz_3 + d} = 1\eqno(4.5)$$
where $\zeta$  will be taken to infinity at the end of the calculation. 
Since $ad-bc=1$, this set of equations can be solved for the parameters 
$a, b, c, d$ and the solutions are given by 
$$\pmatrix{
a & b\cr
c& d\cr} = \pmatrix {\sqrt {{\zeta z_{31}\over (\zeta -1)z_{23}z_{12}}}& 
 -z_2\sqrt {\zeta z_{31}\over (\zeta - 1)z_{23}z_{12}}\cr
- {\zeta z_{23} + z_{12}\over \sqrt {\zeta (\zeta - 1)z_{12}z_{23}
z_{31}}} &
{\zeta z_1 z_{23} + z_3 z_{12}\over\sqrt{\zeta (\zeta - 1)z_{12}
z_{23} z_{31}}}\cr } 
\eqno (4.6)$$ 
Using these equations it follows that 
$$cz_I + d = {z_{12}z_{3I} + \zeta z_{23}z_{1I}\over 
\sqrt {\zeta (\zeta -1)z_{12}z_{23}z_{31}}}  \eqno(4.7) $$
The  integral $I_N$ then becomes 
$$\eqalign {I_N&= \prod_{I =1}^N \Bigl\lbrack\mid \sqrt {{\zeta z_{23}\over 
(\zeta - 1) z_{12}z_{31}}} z_{1I} + \sqrt {{ z_{12}\over \zeta (\zeta - 1)
z_{23}z_{31}}} z_{3I} \mid\Bigr\rbrack^{4(q-\Sigma )\alpha_I} \cr& ~~~\times
\int dK(x_i, y_r)\prod_i\prod_r \prod_{K = 4}^N \mid x_i - r_K
\mid^{-4b\alpha_K} \mid y_r - r_K\mid^{-4c \alpha_K}\cr 
 &~~~\times\mid x_i - 
\zeta\mid^{-4b\alpha_1} \mid x_i\mid^{-4b\alpha_2} \mid x_i-1\mid^{-4b\alpha_3}
\mid y_r - \zeta\mid^{-4c\alpha_1} \mid y_r\mid^{-4c\alpha_2} \mid y_r-1
\mid^{-4c\alpha_3}\cr}\eqno(4.8)$$ 
where the $r_K$ are given by 
$$r_K = {\zeta z_{K2}z_{31}\over z_{12}z_{3K} + \zeta z_{23}z_{1K}},~~~~~K\geq 4
\eqno(4.9)$$  
On taking the limit $\zeta\rightarrow \infty$, $r_K$ become the conformally
invariant cross-ratios 
$$r_K={ z_{K2}z_{31}\over z_{K1}z_{32} }, ~~~~~~ K\geq 4 \eqno(4.10)$$
and 
$$\eqalign{
I_N &= \mid {z_{12}z_{31}\over z_{23}}\mid^{-2(q - \Sigma )\alpha_1}  
\prod_{I = 2}^N\mid {z_{23}\over z_{12}z_{31}}\mid^{2(q-\Sigma )\alpha_I} 
\mid z_{1I} \mid^{4(q-\Sigma )\alpha_I}\cr& 
\times \int dK(x_i, y_r) \prod_i \prod_r \prod_{K = 4}^N \mid x_i - r_K
\mid^{-4b \alpha_K}\mid y_r - r_K \mid^{-4c \alpha_K}\cr
&~~~\times \mid x_i\mid^{-4b\alpha_2}\mid 
x_i-1\mid^{-4b\alpha_3} \mid y_r\mid^{-4c\alpha_2}\mid y_r-1\mid^{-4c\alpha_3}
\cr} \eqno(4.11)$$ 
where the $\zeta$-dependent terms from the integrand cancel those coming from 
the prefactor of (4.8). In the above computation we have chosen to extract
$z_1,~ z_2$ and $z_3$. But of course, since the $N$-point function is 
completely symmetric under the permutation of the indices $I = 1\cdots N$, we
could have used any three $z_I$ for the SL (2, C) transformations. Thus (4.11)
is invariant under these permutations, although it is not manifestly so because
the cross-ratios change with the permutations. This invariance is what is
usually referred to as crossing symmetry. Note that it gives relations between
correlation functions of the same order $N$, but does not connect correlation 
functions of different order. By using a standard relation between four and 
three point functions in conformal field theories, the four-point crossing 
relations can be used to obtain constraints on the coefficient of the 
three-point function. An indirect method of checking that the DOZZ proposal 
satisfies these constraints was used in [4]. In the next section we shall 
directly derive the three-point function from the results of this section.      
\bigskip
\centerline {{\bf V. THE THREE-POINT FUNCTION}}
\bigskip
The simplest case of (4.11) is the case of the three-point function which is
distinguished by the fact that there are no cross-ratios. In that case, (4.11)
reduces to 
$$I_3 (b, c; \alpha_1 , \alpha_2 , \alpha_3; m, n) =  \mid {z_{12}z_{31}\over
z_{23}}\mid^{-2(q - \Sigma )\alpha_1} \mid {z_{12}z_{23}\over z_{31}}
\mid^{-2(q-\Sigma )\alpha_2} \mid {z_{23}z_{31}\over z_{12}}\mid^{-2(q-\Sigma )
\alpha_3}{\cal I}_3\eqno(5.1)$$ 
where ${\cal I}_3$ is defined by 
$${\cal I}_3 (m, n) = \int dK(x_i, y_r) \prod_i \prod_r  
\mid x_i\mid^{-4b\alpha_2} \mid x_i -1\mid^{-4b\alpha_3}\mid y_r
\mid^{-4c \alpha_2}\mid y_r - 1\mid^{-4c\alpha_3}\eqno(5.2)$$ 
Denoting the extrapolation of the above integral from integer values of $m$ and
$n$, by ${\cal I}_3(iu, iv)$ and substituting in (3.11) and (3.6), we get
for the three-point function  
$${\cal G}_3 (z_1, z_2, z_3;\alpha_1,\alpha_2,\alpha_3) =  
{\cal H}_3\mid z_{12}\mid^{2(\Delta_3 - \Delta_1 - \Delta_2 )}
\mid z_{23}\mid^{2(\Delta_1 - \Delta_3 - \Delta_2 )}
\mid z_{31}\mid^{2(\Delta_2 - \Delta_3 - \Delta_1 )}\eqno(5.3)$$
where  
$${\cal H}_3 (\alpha_1 , \alpha_2 , \alpha_3 )= -{1\over 8}\int dudv~{
{\cal I}_3(iu, iv)\over \Gamma (1 + iu)\Gamma (1 + iv )} {\delta (bu + cv - 
\xi )\over {\hbox{sinh}}\pi u{\hbox {sinh}}\pi v}\Bigl({\mu_b^R\over\mu}
\Bigr)^{iu}\Bigl( {\mu_c^R\over \mu}\Bigr)^{iv} \eqno(5.4)$$
We shall now discuss how the extrapolation is done. Note that the final result 
for the three-point function is got by putting together the contributions from 
the zero mode integral and the fluctuation mode integral. As already 
mentioned, the latter is known only at a discrete set of points labelled by 
positive integer values. The former, however, is a known function of $\xi$, 
for negative values of $\xi$. Therefore we continue the zero mode integral to 
positive values of $\xi$ where the two contributions can be put together. Once 
this is done, we shall show that it is possible to continue the  full result 
away from the positive integral values to which the fluctuation mode integral 
is restricted.  

Using the computations of Dotsenko and Fateev [7] and the $k$-function 
introduced by the Zamolodchikovs (defined in B8), we first note, as shown in 
Appendix B, that for positive integer $m$ and $n$ the fluctuation mode 
contribution (5.2) can be written as  
$${\cal I}_3 (m,n) = -m!n! \Phi_b^m \Phi_c^n {k'(0)\over k'(-bm - cn)}
\prod_{I = 1}^3 {k(2\alpha_I)\over k(\Sigma - 2\alpha_I)} \eqno(5.5)$$
where 
$$\Phi_b = - \pi {bk(2b)\over k(b)}\eqno(5.6)$$
Here the function $k(x)$ is an entire function with simple zeroes at 
$x = - (mb + nc)$ and $x = (m + 1)b + (n + 1)c$ for all $m,n \geq 0$.
It has the reflection symmetry property $k(x) = k (q -x)$ and is quasi-periodic 
in the sense that 
$$k (x + b) = k (x)\gamma (bx) b^{1-2bx},~~~ 
k (x + c) = k (x)\gamma (cx) c^{1-2cx},~~~\gamma (x ) = {\Gamma (x) \over 
\Gamma (1 - x)}\eqno(5.7)$$ 
The terms $m!n!\Phi_b^m\Phi_c^n$ and the $\alpha_I$ parts of this expression 
have a straightforward extrapolation from $m, n$ to $u, v$ but, as we shall see,
the extrapolation of the function $k'(-mb - nc)$ is somewhat ambiguous. So for 
the moment, we simply assume that it has an extrapolation to some function 
$\tilde k (-bu - cv)$. The extrapolation of (5.5) is then  
$${\cal I}_3 (iu, iv) = -\Gamma (1 + iu) \Gamma (1 + iv) 
\Phi_b^{iu}\Phi_c^{iv}{k'(0)\over \tilde k(-bu - cv)} \prod_{I = 1}^3 
{k(2\alpha_I)\over k(\Sigma - 2\alpha_I)} \eqno(5.8)$$
Substituting this result in the expression for ${\cal H}_3$, the $\Gamma$ 
functions exactly cancel and we obtain 
$${\cal H}_3 = {1\over 8}\int du dv~{\delta (bu + cv - \xi )\over
{\hbox{sinh}} \pi u{\hbox {sinh}}\pi v} B^{iu} C^{iv} {k'(0)\over \tilde 
k(-bu - cv)}\prod_{I = 1}^3{k(2\alpha_I)\over k(\Sigma - 2\alpha_I)}\eqno(5.9)$$
where 
$$B = {\mu_b^R\over\mu}\Phi_b ,~~C = {\mu_c^R\over\mu}\Phi_c  \eqno(5.10)$$
We may rewrite (5.9) in the form 
$${\cal H}_3 = {1\over 8} L(\xi ) {k'(0)\over \tilde k(-\xi )} 
\prod_{I = 1}^3 {k(2\alpha_I)\over k(\Sigma - 2\alpha_I)} \eqno(5.11)$$
where 
$$L(\xi ) =   \int du dv {\delta (bu + cv - \xi )
\over{\hbox{sinh}} \pi u{\hbox{sinh}}\pi v}B^{iu}C^{iv} \eqno(5.12)$$
In arriving at the above equations we have used the fact that the delta 
function converts the functions of $bu + cv$ into functions of $\xi$ which 
enabled us to remove all the $k$-dependent parts from the $u, v$ integrations.

The function $L (\xi)$ is closely related to the function ${k'(\xi )\over k(
\xi )}$ of the Zamolodchikovs. To see this we note that the $L(\xi )$ function 
satisfies the following recursion relations:
$$L(\xi - ib) = -B^{-1}\Bigl\lbrack L(\xi ) - 2{C^{ib\xi}\over{\hbox {sinh}}
\pi b \xi}\Bigr\rbrack\eqno(5.13)$$
and a similar one for $\xi \rightarrow \xi + c$. The above equation may be 
compared with the analogous relations we can derive for the function 
${k'(\xi )\over k(\xi )}$ which have, instead of ${1\over{\hbox{sinh}}\pi b
\xi}$ on the right hand side ${\gamma '(b\xi )\over\gamma (b\xi )}$, and 
similarly for $\xi \rightarrow \xi + c$. 
Thus, both functions have simple poles at the points $\xi = \pm (mb + nc)$, but
while $L(\xi )$ has residues $B^mC^n$, ${k'(\xi )\over k (\xi )}$ has residues 
equal to $1$. Thus  
$$\lim_{\xi \rightarrow mb + nc} L(\xi )k(\xi ) = B^mC^n k' (mb + nc)
~~~~m,n \geq 0\eqno(5.14)$$ 
Substituting this back in (5.11) we find that the coefficient of the
three-point function may be written as  
$${\cal H}_3 (\alpha_1, \alpha_2, \alpha_3) = {1\over 8}\Bigl\lbrack L(-\xi )
{k(-\xi ) \over k'(-\xi )}\Bigr\rbrack\times {1\over k(-\xi )}\prod_{I = 1}^3 
{k(2\alpha_I)\over k(\Sigma - 2\alpha_I)} \eqno(5.15)$$
Note that the factor in the parantheses on the right hand side of the above 
equation is finite. This factor in the two-exponential theory is more general 
than the corresponding factor in the one-exponential theory. If this is taken 
into account, an analysis similar to the one in [4] would probably reveal that 
the two-exponential theory is also crossing-symmetric since the rest of the 
expression is readily seen to agree with the conjecture of the Zamalodchikovs 
in [5]. It is however interesting to point out that the expression in (5.15) 
has a rather complicated behaviour under a reflection $\alpha\rightarrow q - 
\alpha$ because of the more general factor in the parantheses. 

Since the dimensionless parameters ${\mu_b^R\over\mu}$ and ${\mu_c^R\over\mu}$ 
are at our disposal, we may now specialise to the case which simplifies the 
$u,v$ integrations in $L(\xi )$ by letting 
$$B^c = C^b \equiv A ~~~ {\hbox {which implies}}~~~ \bigl(-\pi{\mu_b^R\over\mu}
\gamma (b^2)\bigr)^c = \bigl(-\pi{\mu_c^R\over\mu}\gamma (c^2)\bigr)^b 
\eqno(5.16)$$ 
Denoting this simplified integral by $l(\xi )$ we find 
$$l (\xi ) = \int du dv~ {\delta (bu + cv - \xi )\over{\hbox{sinh}}\pi 
u{\hbox {sinh}}\pi v}\eqno(5.17)$$ 
The function $l(\xi )$ has residues $\pm 1$ at the positive and negative 
poles respectively while the function ${k'(\xi )\over k(\xi )}$ has 
residues 1 at all poles.  This corresponds to the fact that $l (\xi)$ is  
even,  and ${k'(\xi )\over k(\xi )}$ is odd under the reflection transformation 
$\xi \rightarrow q - \xi$. The important point however is that the   
residues coincide for positive $m$ and $n$ and thus  
$$\lim_{\xi \rightarrow mb + nc}l(\xi )k(\xi ) =  k' (mb + nc)
~~~~m,n \geq 0\eqno(5.18)$$ 
This shows that there are at least two natural choices for  $\tilde k (\xi)$
as an extension of $k' (mb + nc),~~~m,n\geq 0$ namely, $\tilde k (\xi ) 
= k' (\xi )$ (odd) and $\tilde k (\xi ) = l(\xi )k (\xi )$ (even). 
Any arbitrary combination of the above two choices is also permissible. 
And there may be other possibilities. However, if we now require that the 
extension $\tilde k (\xi )$ does not introduce any new singularities 
(i.e. not already occuring in $l(\xi ))$, then we see that  
$${l(\xi )\over \tilde k (\xi )} k(\xi )  = e (\xi ) \eqno(5.19)$$
where from (5.18), $e(\xi )$ is an entire function satisfying 
$$e (mb + nc ) = 1, ~~~m,n \geq 0 \eqno(5.20)$$   
But this means that $e(\xi )$ is doubly periodic on the positive real axis, 
which by Jacobi's theorem [10] on doubly periodic functions means that $e = 1$ 
everywhere. Thus the only choice of $\tilde k$ that does not introduce new 
poles is 
$$\tilde k(\xi ) = l(\xi)k(\xi) \eqno(5.21)$$
In that case 
$${\cal H}_3 = {1\over 8}{k'(0)\over k(-\xi)}A^\xi\prod_{I = 1}^3
{k(2\alpha_I ) \over k(\xi + 2\alpha_I)}\eqno(5.22)$$
This expression is covariant under a reflection $\alpha\rightarrow q - \alpha$
because the denominator is invariant under this transformation and the only
non-trivial contributions come from the behaviour of the numerator.  
Substituting the above equation in (5.3) we get finally 
$${\cal G}_3 = {1\over 8}{k'(0)\over k(-\xi)}A^\xi \prod_{I = 1}^3 
{k(2\alpha_I )\over k(\xi + 2\alpha_I)}\mid z_{12} \mid^{2(\Delta_3 - 
\Delta_1 - \Delta_2 )}\mid z_{23}\mid^{2(\Delta_1 - \Delta_3 - \Delta_2 )}
\mid z_{31}\mid^{2(\Delta_2 - \Delta_3 - \Delta_1 )} \eqno(5.23)$$
Note that this expression is exactly the one that was conjectured by the 
Zamolodchikovs on the basis of the one-potential theory.  Thus we have shown 
that this result can be derived in the two-potential theory in a natural way.
Furthermore, whereas in the one-potential theory, only one set of poles could
be physically identified; in the two-potential theory, the full lattice of 
poles can be identified. 
\bigskip
\centerline {\bf {VI. THE TWO-POINT FUNCTION}}  
\bigskip
The two-point function can be evaluated directly along the lines of the 
three-point function. It is easy to see that the translational invariance 
and scale covariance properties of the path integral imply that the fluctuating
part has the following structure 
$$I_2(b, c; \alpha_1,\alpha_2; m, n) = ~\mid z_{12}\mid^{2(q - \Sigma )\Sigma}
{\cal I}_2\eqno(6.1)$$ 
where 
$${\cal I}_2 =  \int dK(x_i, y_r)\prod_i\prod_r
\mid x_i - 1\mid^{-4b\alpha_1} \mid y_r - 1\mid^{-4c\alpha_1}
\mid x_i\mid^{-4b\alpha_2}\mid y_r\mid^{-4c\alpha_2} \eqno(6.2)$$
Substituting for $I_2$ from the above equations in (3.11) and (3.6), we get 
$${\cal G}_2 (z_1, z_2, \alpha_1, \alpha_2 ) = {\cal I}_2 (b, c; 
\alpha_1,\alpha_2; m,n)\mid z_{12}\mid^{-2(\Delta_1 + \Delta_2 )}\eqno(6.3)$$
The integral in (6.2) is essentially the same as the one encountered in the 
computation of the three-point function. But, in contrast to the latter, 
it is infinite or zero according as $D = \Delta_1 - \Delta_2 =  0$ or not. 
This is due to the SL(2, C) covariance and can be seen as follows: Under any 
SL(2, C) transformation with parameters $\{a, b, c, d\}$, the integral and 
the quantity $\mid z_1 - z_2\mid^{-1}$ pick up the usual factors 
$$\mid cz_1 + d\mid^{-2\Delta_1} \mid cz_2 + d\mid^{-2\Delta_2}
~~~\hbox {and}~~~ 
\mid cz_1 + d\mid \mid cz_2 + d\mid \eqno(6.4)$$
respectively. Consider now the stability subgroup $S$ of $z_1$ and $z_2$. 
This is a one-parameter subgroup whose standard parameters 
$\{a_s, b_s, c_s, d_s\}$ say, are functions of one free parameter. Since the 
transformations belonging to the subgroup leave both the integral and the 
quantity $\mid z_1 - z_2\mid$ invariant, we have from (6.4)  
$$ I_2 = \mid c_sz_1 + d_s\mid^{-2\Delta_1} \mid c_sz_2 + d_s\mid^{-2\Delta_2} 
I_2 ~~~{\hbox{and}}~~~ \mid c_sz_1 + d_s\mid \mid c_sz_2 + d_s\mid = 1 
\eqno(6.5)$$
and thus 
$${\cal I}_2 (\xi , \eta ) = (c_sz_1 + d_s)^D{\cal I}_2 (\xi, \eta )~~~{\hbox 
{where }}~~~\xi = q - \alpha_1 -\alpha_2 ,~\eta = \alpha_1 - \alpha_2,~~{\hbox 
{and}}~~D = \xi\eta \eqno(6.6)$$
which shows that either $I_2 = 0$ or $D = 0$. In the case that $D= 0$, 
the integral is invariant under a change of variable corresponding to the 
subgroup and is therefore infinite. To obtain a finite result, the integration 
corresponding to the one-parameter subgroup must be factored out. 

Since $D = (q  - \alpha_1 - \alpha_2 )(\alpha_1 - \alpha_2 ) $ we see that 
$D = 0$ corresponds to either $\alpha_1 = \alpha_2$ or $\alpha_1 = q - 
\alpha_2$ {\it i.e.} the parameters $\alpha_1$ and $\alpha_2$ are either equal  
or reflection conjugate. The fact that the two-point function is zero except 
for these values is not surprising when we recall that the two-point function 
may be regarded as the inner product of two formal states of the kind 
$\mid\alpha , z> = e^{\alpha\phi (z)}\mid 0>$ and these would be expected to 
be orthogonal unless the parameters were conjugate in some sense.  However, as 
already mentioned, the integral is infinite and has to be regulated by 
factoring out the one-parameter subgroup $S$. A natural way to regulate it is
to note from (5.2) and (6.2) that the two-point function is the $\alpha_3 
\rightarrow 0$ limit of the three-point function with coefficient    
$$ l(\xi - \alpha_3) I_3 (\alpha_1, \alpha_2, \alpha_3) = 
{k'(0)k(2\alpha_1)k(2\alpha_2)k(2\alpha_3)\over k(\alpha_3 - \xi )
k(\alpha_3 + \xi )k(\alpha_3 - \eta )k(\alpha_3 + \eta ) }\eqno(6.7)$$ 
Since by definition $D = \xi\eta$, we see that $D\neq 0$ implies $\xi \neq 0$
and $\eta\neq 0$ and from (6.7) we see that in this case the function does 
indeed vanish as $\alpha_3 \rightarrow 0$. On the other hand $D = 0$ implies 
$\xi = 0$ or $\eta = 0$, in which cases (6.7) becomes     
$${k'(0)k(2\alpha_1)k(2\alpha_2)k(2\alpha_3)\over k(\alpha_3 )
k(\alpha_3 )k(\alpha_3 - \eta )k(\alpha_3 + \eta )} 
~~~{\hbox {or}}~~~ {k'(0)k(2\alpha_1)k(2\alpha_2)k(2\alpha_3)\over 
k(\alpha_3 - \xi ) k(\alpha_3  + \xi )k(\alpha_3 )k(\alpha_3 )} \eqno(6.8)$$ 
respectively. A short computation shows that in the limit $\alpha_3 
\rightarrow 0$ these become  
$$ {2\over \alpha_3}~~~{\hbox{and}}~~~{2\over\alpha_3}{k(\xi )\over k(-\xi )}
\eqno(6.9)$$ 
Interpreting the universal constant $2/\alpha_3$ as the integral over the 
stability subgroup that has to be factored out, we obtain finally 
$$G_2 (q - \alpha , \alpha ; z_1, z_2) = G_2 (\alpha , q - \alpha ; z_1, z_2) 
= \mid z_{12}\mid^{-4\Delta}\eqno(6.10)$$ 
$$G_2 (\alpha , \alpha ; z_1, z_2) = G_2 (q - \alpha , q - \alpha ; z_1, z_2) 
= N(\alpha )\mid z_{12}\mid^{-4\Delta}\eqno(6.11)$$ 
where 
$$N(\alpha ) = \Bigl\lbrack {k(\xi )\over k(-\xi )}\Bigr\rbrack_{\xi = q - 
2\alpha} = {k(2\alpha )\over k(2\alpha - q)}\eqno(6.12)$$ 
The ambiguity in the two-point function may indicate that the extrapolated 
two-point function (and hence the one-point and partition functions) do not 
really exist {\it i.e.} that the functional integral is a kind of distribution
which takes meaningful values only when tested against products of at least 
three external fields. 
\bigskip
\centerline {\bf {VII. THE FOUR-POINT FUNCTION}}  
\bigskip
The four-point function may be calculated along the same lines as the  
three-point function. It is straightforward to see that it takes the form  
$$ \eqalign{ {\cal G}_4 (z_1, z_2, z_3, z_4;\alpha_1,\alpha_2,\alpha_3,\alpha_4)
&=  P(z_{IJ})\times Q(r, \bar r)\cr&\times \int du dv~{\delta 
(bu + cv - \xi )\over{\hbox {sinh}}u{\hbox {sinh}}v}I(r, \bar r, u, v)}
\eqno(7.1)$$
where 
$$ P (z_{IJ}) = \mid z_{12}\mid^{-2(\Delta_1 + \Delta_2 - \Delta_3 - \Delta_4)}
\mid z_{23}\mid^{2(-\Delta_1 + \Delta_2 + \Delta_3 + \Delta_4)}\mid 
z_{13}\mid^{2(\Delta_1 - \Delta_2 + \Delta_3 - \Delta_4)}\mid z_{14} \mid^{ 
4\Delta_4}\eqno(7.2)$$
$$ Q(r, {\bar r}) = \mid r\mid^{4\alpha_2\alpha_4}
\mid r - 1\mid^{4\alpha_3\alpha_4}\eqno(7.3)  $$
and $I(r, \bar r, m, n)$ is given by  
$$\eqalign {I (r, &\bar r, m, n) = \int dK(x_i, y_r)\prod_i\prod_r
\mid x_i - r\mid^{-4b\alpha_4} \mid y_r - r\mid^{-4c\alpha_4}
\cr&\times \mid x_i\mid^{-4b\alpha_2}\mid y_r\mid^{-4c\alpha_2}\mid x_i - 1
\mid^{-4b\alpha_3} \mid y_r - 1\mid^{-4c\alpha_3} }\eqno(7.4)$$ 
In the special case when both $-4b\alpha_4$  and $-4c\alpha_4$ are positive 
integers (which implies that $b$ is rational), the integral in (7.4) can be 
expanded as a polynomial in $r$ with coefficients which are the three-point 
structure constants. Although $\alpha_4$ is singled out here,  
it is clear that as long as one of the  exponents is a positive 
integer, by choosing an appropriate SL(2, C) transformation, the four-point 
function is a polynomial in a suitable cross-ratio. 

In the generic case, where none of the $\alpha$s is a positive integer, if one 
divides the integral (7.4) into sections for which 
$x_i$ and $y_r$ are greater or less than $r$, one can expand each of these 
sections in an infinite power series in $r$ or $r^{-1}$. 
The simplest example is for $m = 1,~ n =0$ when we have an integral of the form
$$\int d^2x~ \mid x\mid^{-4b\alpha_2} \mid x - 1\mid^{-4b\alpha_3}\mid x - r
\mid^{-4b\alpha_4}\eqno(7.5)$$ 
These integrals can be split into parts $I_>$ and $I_<$ for which $\mid x\mid >
\mid r\mid$ and $\mid x\mid <\mid r\mid$ respectively such that 
$$I(r, \bar r) = I_> + I_<\eqno(7.6)$$
$I_>$ and $I_<$ have the expansions     
$$\eqalign{I_> (r,\bar r) &= \sum_{p = 0}^\infty\int d^2x~\mid x\mid^{-4b
\alpha_2}\mid x - 1 \mid^{-4b \alpha_3} \Bigl({-4b\alpha_4\atop p}\Bigr)
\mid r\mid^p\mid x \mid^{-p}\mid x \mid^{-4b \alpha_4}\cr& = \sum_{p = 
0}^\infty \Bigl({-4b\alpha_4\atop p}\Bigr)\mid r\mid^p\int d^2x \mid x
\mid^{-4b(\alpha_2 + \alpha_4 + {p\over 4b})}\mid x - 1\mid^{-4b\alpha_3}\cr
&\sum_{p = 0}^\infty \Bigl({-4b\alpha_4\atop p}\Bigr)
{\cal H}_3 (-\alpha_2 -\alpha_4-\alpha_3 - {p\over 4b} + c, \alpha_2 + 
\alpha_4 + {p\over 4b}, \alpha_3 )\mid r \mid^p} \eqno(7.7a)$$ 
and 
$$\eqalign{I_< (r,\bar r) &= \sum_{p = 0}^\infty\int d^2x~\mid x\mid^{-4b
\alpha_2 + p}\mid x - 1 \mid^{-4b \alpha_3} \Bigl({-4b\alpha_4\atop p}\Bigr)
\mid r\mid^{-p}\cr& = \sum_{p = 0}^\infty\Bigl({-4b\alpha_4\atop p}\Bigr)\mid r
\mid^{-p}\int d^2x \mid x\mid^{-4b(\alpha_2 - {p\over 4b})}\mid x - 1
\mid^{-4b\alpha_3}\cr &\sum_{p = 0}^\infty \Bigl({-4b\alpha_4\atop p}\Bigr)
{\cal H}_3 (-\alpha_2 -\alpha_3 + {p\over 4b} + c, \alpha_2 - {p\over 4b}, 
\alpha_3 )\mid r \mid^{-p}} \eqno(7.7b)$$ 
respectively where the coefficients ${\cal H}_3$ are just the three-point 
coefficients already discussed in Section V. Note that each of these expansions
is only an asymptotic expansion of the full integral. 
\bigskip
\centerline {{\bf VIII. CONCLUSIONS}}
\bigskip
In this paper we have considered the quantisation of the two-dimensional 
Liouville field theory by computing the N-point functions of vertex operators 
using path integral methods. It is argued that the 
standard one exponential Liouville potential admits a two-exponential
generalisation because, in the quantum theory, there are two fields, rather 
than one, with a given conformal weight. It is shown that the two-exponential  
theory is not only conformally invariant but also has a built-in duality 
symmetry which was inferred earlier from the form of the N-point functions 
of the standard Liouville theory.  
We have derived expressions for the N-point functions and explicitly 
computed the three-point function. 
We have shown that the coefficient of the three-point function 
exhibits a two-dimensional lattice of poles in the parameter space.  
Unlike previous work where the existence of the two-dimensional 
lattice (as opposed to a one-dimensional lattice) was inferred in an indirect 
manner, the existence
of the two-dimensional lattice is shown to be a natural and direct consequence 
of the quantum mechanical duality symmetry of the Liouville theory.  

Unlike the three-point function, the lower point functions are invariant under 
non-compact sub-groups of SL(2, C) and are therefore infinite. In principle 
they can be made finite by factoring out the integral over the relevant 
subgroups. We propose a method of doing this for the two-point function 
and show that it vanishes unless the two vertex parameters are either 
equal or reflection conjugate. The four-point function can also be studied 
using our methods. We have briefly discussed its most important properties. 
In particular we have shown that when one of the 
vertex functions has an integer power, the conformal block of the four-point 
function is given by a known polynomial of the cross-ratio. In general,
however, the conformal block can only be obtained as a sum of asymptotic 
expansions in the cross-ratio and its inverse.     

We conclude by suggesting some generalisations.  
It is probable that the main aspects of the   
formalism we have presented are equally valid for Toda 
theories and for their supersymmetric generalisations. 
It is also clear from our analysis that  much of the calculation is 
valid for  
any two-exponential potential theory. It would be therefore interesting to see 
how far we can extend these results to interesting integrable field theories  
like the Sine-Gordon and Sinh-Gordon theories. 
We hope to generalise our analysis to these theories in the future. 
\bigskip
\vfil\eject
\centerline {{\bf Appendix A: The Finite-Volume Green's Function}}
\bigskip
The Green's function $G(x,y)$ on a compact two-dimensional space $S^2$ of 
volume $\Omega$ is the inverse of $\sqrt{g}\Delta_x$ where $\Delta_x$ is the 
two-dimensional Laplace-Beltrami operator defined in (2.2). Since on a 
two-dimensional compact space the only zero mode of $\Delta_x$ is the 
constant function it follows that $G(x,y)$ is the (unique) solution of the
equation
$$\Delta_x G(x,y)={\pi \over \sqrt{g}}\delta^2(x-y)~~~ \hbox{on}~~~ 
L_2(S)\ominus P_0\eqno(A1)$$
which is orthogonal to the zero modes 
$$\int d^2z\sqrt{g(z)}G(z,y)=\int d^2y\sqrt{g(y)}G(z,y)=0 \eqno(A2)$$
the expression $L_2(S)\ominus P_0$ denoting the usual Hilbert space for $S$ 
minus the projection $P_0$ on the zero modes. It is not difficult to verify
that in conformal coordinates the solution of (A1) is
$$ G(x,y) = G_0(x,y) - {1 \over \Omega} [\rho(x)+\rho(y)] + {1\over \Omega^2}
\int d^2z\sqrt{g(z)}\rho(z) \eqno(A3)$$
where 
$$G_0(x,y) = -{1\over 2}\hbox{ln}{\mid x-y\mid \over L} \eqno(A4)$$
$L$ is the renormalization scale, and 
$$\rho(x)=\int d^2y\sqrt{g(y)}G_0(x,y) \eqno(A5)$$
Clearly it is the $\rho$ terms in (A3) that make $G(x,y)$ orthogonal to the 
zero modes. 

It is of special interest to study the short-distance behaviour of $G_0(x,y)$ 
i.e. the limit $x\rightarrow y$, because we shall have to define $G(x,y)$ for
coincident points. In conformal coordinates the line-element is given by 
$$ds^2=\sqrt{g(y)}dyd\bar y\eqno(A6)$$
and hence from (A4)  
$$\Bigl(G_0(x,y)\Bigr)_{x\rightarrow y} \rightarrow -{1 \over 2}\hbox{ln}
{\mid dy\mid \over L} = - {1 \over 2}\hbox{ln}\bigl( {ds \over L}\bigr)  
 + {1 \over 4}\hbox{ln}(\sqrt{g}) \eqno(A7)$$
where $ds$ is the geodesic distance between $x$ and $y$. It is clear that as
$x\rightarrow y$ the right-hand side of (A7) diverges, but the point is that 
the divergent part contains only $ds$, which is a diffeomorphic 
{\it invariant}. This means that we can absorb the divergent part of $G$ in 
the renormalisation scale $L$, in a diffeomorphic invariant manner, leaving 
only the ${\hbox {ln}}\sqrt{g}$ term. We then interpret the renormalised 
version $G^R_0(x,x)$ of $G_0(x,x)$ as  
$$G^R_0(y,y) = {1\over 4}\hbox{ln}(\sqrt{g(y)}) \eqno(A8)$$
An important point to note is that, although $G_0(x,y)$ is invariant with 
respect to the Weyl transformations $\sqrt{g(x)}\rightarrow \lambda(x)
\sqrt{g(x)}$ the renormalised quantity $G^R_0(x,x)$ is not. In fact it has the 
Weyl transformation $$G_0^R(x,x) ~~~ \rightarrow ~~~ G_0^R(x,x) +  { 1\over 4}
\hbox{ln}(\lambda(x))\eqno(A9)$$
It is also worth noting that, whereas $G_0(x,y)$ is invariant with respect to 
all (rigid and local) Weyl transformations, the functions $\rho(x)$ in (A5) are
invariant only with respect to rigid Weyl transformations. Hence the full
Green's function $G(x,y)$ is invariant with respect to rigid Weyl 
transformations for all $\Omega$ but is invariant with respect to all Weyl
transformations only in the infinite-volume limit. 
\bigskip
\vfil\eject
\centerline {{\bf Appendix B: The Dotsenko-Fateev Integral}}
\bigskip
The integral 
$$\eqalign{ I = &\prod^m_{i<j}\prod^n_{r<s}\int d^2x_id^2y_r\mid x_i-x_j
\mid^{4\rho} \mid y_r-y_s\mid^{4\rho '} \mid x_i-y_r\mid^{-4}\cr&
~~~\times\mid x_i\mid^{2\alpha}\mid x_i - 1\mid^{2\beta}\mid y_r
\mid^{2\alpha'}\mid y_r - 1 \mid^{2\beta'} } \eqno(B1)$$
which depends, apart from $\rho$ and $\rho '$, on the six parameters $m, n, 
\alpha , \beta , \alpha', \beta'$ has been computed by Dotsenko and Fateev [7].
By using the dictionary 
$$\rho = -b^2,~~~ \rho' = -c^2,~~~\alpha = -2b\alpha_1,~~~\beta = -2b\alpha_2,
~~~\alpha ' = -2c\alpha_1,~~~ \beta ' = -2c\alpha_2 \eqno(B2)$$
and the definition 
$$\gamma(x) = {\Gamma(x)\over\Gamma (1-x)}\eqno(B3)$$ 
their result for (5.2) may be written in the form 
$$I = \Bigl[m!n!\pi^{m + n} b^{-8mn}\bigl(\gamma(-b^2)\bigr)^{-m} \bigl(\gamma 
(-c^2)\bigr)^{-n} \Bigr](XYZ)^{-1}\eqno(B4)$$
where 
$$X = \prod_{I = 1}^3\prod_{l = 0}^{m-1}\gamma (2b\alpha_I + lb^2),~~~
Y = \prod_{I = 1}^3\prod_{l = 0}^{n-1}\gamma (2c\alpha_I + m + lc^2)\eqno(B5)$$ 
$$Z = \prod_{l= 1}^m\gamma ( 1 + lb^2)\prod_{l = 1}^n\gamma (1 + m + lc^2)
\eqno(B6)$$
and $\alpha_3$ is an auxiliary variable defined by 
$$\alpha_3\equiv q-(\alpha_1+\alpha_2)-mb-nc \eqno(B7)$$
The problem with this expression, from our point of view, is that it is not in  
a form that readily admits an extrapolation to non-integer values of $m$ and 
$n$. This situation can be remedied by using the function $k(x)$ defined by the 
Zamolodchikovs as 
$$\hbox{ln}(k(x)) = \int_0^{\infty}{dt \over t}\Bigl(({q \over 2} - x)^2e^{-2t}
- {\hbox{sinh}^2({q \over 2} -x)t\over \hbox{sinh}(bt)\hbox{sinh}(ct)}\Bigr) 
\eqno(B8)$$
in the range $0 < x < q$, and elsewhere by analytic continuation. As already 
mentioned in the Section V, the 
relevant properties of $k(x)$ are that it has the symmetry property 
$k(x)=k(q-x)$, it is an entire function with zeros at $x= -mb -nc$ and 
$x = (m+1)b + (n + 1)c$ for any non-negative integers $m$ and $n$ and that it 
is related to $\gamma(x)$ by the recursion relation 
$$\gamma(bx) = {k(x+b)\over k(x)}b^{(2bx-1)} \eqno(B9)$$
>From the recursion relation it follows that  
$$\gamma (b\chi + lb^2) = {k\bigl(\chi + (l+1)b\bigr)\over k (\chi + lb)}
b^{(2b\chi + 2lb^2 - 1)} \eqno(B10)$$
and thus 
$$\prod_{l = m_1}^{m_2}\gamma (b\chi + lb^2) = {k\bigl(\chi+(m_2+1)b\bigr)\over 
k(\chi + m_1b)}b^{m[2b\chi - 1 + Mb^2]} \eqno(B11)$$
where $m = m_2 - m_1 + 1$ and $M = m_1 + m_2$. In particular 
$$\prod_{l = 0}^{m-1}\gamma (b\chi + lb^2) = {k(\chi + mb)\over 
k(\chi )}b^{m[(2b\chi - 1) + (m-1)b^2]} \eqno(B12)$$
and 
$$\prod_{l = 1}^m\gamma (b\chi + lb^2) = {k\bigl(\chi + (m+1)b\bigr)\over 
k(\chi + b)}b^{m[(2b\chi-1) + (m+1)b^2]} \eqno(B13)$$
Similar results are valid for $b\rightarrow c, m\rightarrow n$. 
These relations permit us to write the $m$ and $n$ products occurring in $X$, 
$Y$ and $Z$ as ratios of single functions. Thus using (B12) for $X$ and $Y$ we 
obtain 
$$X = \bigl(b\bigr)^{m[4b\Sigma - 3 + 3(m-1)b^2]}\prod_{I = 1}^3
{k(2\alpha_I + mb)\over k(2\alpha_I)} \eqno(B14)$$
and 
$$Y = \Bigl(c\Bigr)^{n[4c\Sigma - 3 + 6m + 3(n-1)c^2]} \prod_{I = 1}^3
{k(2\alpha_I + mb + nc)\over k(2\alpha_I + mb)} \eqno(B15)$$
where $\Sigma = \alpha_1 + \alpha_2 + \alpha_3$. 
A crucial point is that when we combine these expressions to form $XY$ the 
factors $k(2\alpha_I + mb)$ in $X$ and $Y$ cancel to give 
$$ XY = \bigl(b\bigr)^{(mb - nc)\Sigma -6mn} \prod_{I = 1}^3
{k(2\alpha_I + mb + nc)\over k(2\alpha_I)} \eqno(B16)$$
which, using $k(q-x)=k(x)$ and the definition of $\alpha_3$, may be written as 
$$ XY = \bigl(b\bigr)^{(mb - nc)\Sigma - 6mn} \prod_{I = 1}^3 {k(\Sigma - 2
\alpha_I )\over k(2\alpha_I)}~~~ \hbox{where}~~~\Sigma = \alpha_1 + \alpha_2 +
\alpha_3 \eqno(B17)$$
It is this cancellation that is responsible for the equality of the 
two-exponential and one-exponential computations.  
In order to apply the same procedure to $Z$ we have to be a little careful as
the $k$-functions have zeros at relevant points. To allow for this we use the 
formula (B13) with $\epsilon\not= 0$ for $Z$, and take the limit $\epsilon
\rightarrow 0$. We then obtain 
$$Z = \bigl(b\bigr)^\delta\Bigl\{\Bigl[{k\bigl(c + (m + 1)b + \epsilon\bigr)
\over k (c + b + \epsilon )} \Bigr] \Bigl[{k\bigl((1 + m)b + ( n + 1)c + 
\epsilon \bigr)\over k \bigl((1+m)b + c + \epsilon \bigr)}\Bigr]\Bigr\}_{
\epsilon = 0} \eqno(B18)$$
where
$$ \delta = m[1 + (m + 1)b^2] - n[(1 + 2m) + (n + 1)c^2]\eqno(B19)$$
Here again the $k$ factors that contain $m$ but not $n$ cancel, and we obtain 
$$Z = \bigl(b\bigr)^{(mb-nc)(2q-\Sigma)-2nm } \Bigl[{k\bigl((1 + m)b + (n + 1)c
+ \epsilon \bigr)\over k(c + b + \epsilon )}\Bigr]_{\epsilon = 0} \eqno(B20)$$
Note that the terms proportional to $q$ in the exponent $\delta$ add rather 
than cancel as they did for $XY$. This is because the summation runs from $1$ 
to $m$ rather than $0$ to $m-1$ as it did for $X$ and $Y$. Using $k(x)=k(q-x)$
and the definition of $\alpha_3$, this may be written as 
$$\eqalign{ Z &= \bigl(b\bigr)^{(mb - nc)(2q - \Sigma ) - 2nm}\times \Bigl[{k
\bigl( (-mb-nc) - \epsilon \bigr)\over k(q + \epsilon )}\Bigr]_{
\epsilon = 0} \cr &= -\bigl(b\bigr)^{(mb - nc)(2q - \Sigma) - 2nm }
\times {k'(-mb-nc)\over k'(q)}\cr}\eqno(B21)$$
where in the last step we have used L'Hospital's rule. Combining the factors
XYZ we then have 
$$XYZ = -\bigl(b\bigr)^{[2(mb - nc)q - 8mn]}\times {k'(-mb-nc)\over 
k'(0)}\prod_{I = 1}^3 {k(\Sigma - 2\alpha_I)\over k(2\alpha_I )}\eqno(B22)$$
Hence if we define 
$$\Phi_b = - \pi (b^2)^{2 - qb}\gamma(b^2) = -\pi{bk(2b)\over k(b)} \eqno(B23)$$
and use (B4) we have finally 
$$I_{mn} = -m!n! \Phi_b^{m} \Phi_c^{n} {k'(0)\over k'(-mb - nc)}
\prod_{I = 1}^3 {k(2\alpha_I)\over k(\Sigma - 2\alpha_I)} \eqno(B24)$$
\bigskip
\bigskip
{\bf {Acknowledgements}}: This work was begun in collaboration with I. Sachs 
and C. Wiesendanger. The discussion of Weyl invariance in this paper is 
inspired to a large extent by some unpublished work of the latter. It is a 
pleasure to thank J. Teschner for several useful discussions. 
\vfil\eject
\centerline {\bf {REFERENCES}} 
\bigskip
\item {1. } H. Poincare, J. Math. Pure App. {\bf 5} {\it se 4} (1898) 157;
N. Seiberg, Notes on Quantum Liouville Theory and Quantum Gravity, in 
"Random Surfaces and Quantum Gravity", ed. O. Alvarez, E. Marinari, 
and P. Windey, Plenum Press, 1990.
\item {2. } A. Polyakov, Phys. Lett. {\bf B103} (1981) 207; T. Curtright 
and C. Thorn, Phys. Rev. Lett. {\bf 48} (1982) 1309; E. Braaten, T. Curtright 
and C. Thorn, Phys. Lett {\bf 118} (1982) 115; Ann. Phys. {\bf 147} (1983) 365;
J.-L. Gervais and A. Neveu, Nuc. Phys. {\bf B238} (1984) 125; {\bf B238}
(1984) 396; {\bf 257}[FS14] (1985) 59; E. D'Hoker and R. Jackiw, Phys. Rev. 
{\bf D26} (1982) 3517.
\item {3. } L. O'Raifeartaigh and V. V. Sreedhar, Nuc. Phys. {\bf B520}
(1998) 513; Phys. Lett. {\bf B425} (1998) 291. 
\item {4. } J. Teschner, Phys. Lett. {\bf B363} (1995) 65.
\item {5. } M. Goulian and M. Li, Phys. Rev. Lett. {\bf 66} (1991) 2051; 
H. Dorn and H.-J. Otto, Phys. Lett. {\bf B291} (1992) 39;
Nuc. Phys. {\bf B429} (1994) 375; A. and Al. Zamolodchikov, Nuc. Phys. 
{\bf B 477} (1996) 577; Low Dimensional Applications of Quantum Field Theory,
ed. L. Baulieu, V. Kazakov, M. Picco, and P. Windey, NATO ASI Series B,
Physics Volume 362, Plenum Press 1997.     
\item {6. } G. N. Watson, Proc. Roy. Soc. (London) {\bf 95}, (1918) 83;
A. Sommerfeld, "Partial Differential Equations of Physics", Academic Press, 
New York 1949.  
\item {7. } Vl. Dotsenko and V. Fateev, Nuc. Phys. {\bf B251} (1985) 691.
\item {8. } {\it The two-exponential theory was also considered in} 
Vl. S. Dotsenko, Mod. Phys. Lett. {\bf A6} (1991) 3601.
\item {9. } C. Ford and I. Sachs, Phys. Lett. {\bf B418} (1998) 149; 
E. Braaten, T. Curtright and C. Thorn, Ann. Phys. {\bf 147} (1983) 365; 
E. D'Hoker and R. Jackiw, Phys. Rev. {\bf D3517} (1982) 26; H.-J. Otto 
and G. Weigt, Phys. Lett. {\bf B159} (1985) 341; Z. Phys. {\bf C31} 
(1985) 209; J. -L. Gervais and A. Neveu, Nuc. Phys. {\bf B199} (1982) 59;
T. Fujiwara, H. Igarashi and Y. Takimoto, Phys. Lett. {\bf B391} (1997) 78. 
\item {10. } E. T. Copson, An Introduction to the Theory of Functions of 
a Complex Variable, Oxford University Press, London.
\vfil\eject\end